\documentstyle[12pt,aps,pra,epsfig]{revtex}

\begin{document}

\title{\large \bf Ghost Diffraction: Causal Explanation
\\ via Correlated Trajectory Calculations}
\author{Bill Dalton\thanks{email: bdalton@stcloudstate.edu}}
\address{Department of Physics, Astronomy and Engineering Science\\
St. Cloud State University\\
St. Cloud, MN 56301, USA}
\date{\today}
\maketitle

\begin{abstract}

We use trajectory calculations to successfully explain two-photon
"ghost" diffraction, a phenomenon previously explained via quantum
mechanical entanglement. The diffraction patterns are accumulated
one photon pair at a time. The calculations are based on initial
correlation of the trajectories in the crystal source and a
trajectory-wave ordering interaction with a variant generator
inherent in its structure. Details are presented in comparison
with ordinary diffraction calculated with the same trajectory
model.

\end{abstract}

\section{INTRODUCTION}

In the "ghost" diffraction experiment of \cite{Strek}, photons
correlated in a down conversion process are separated and then
counted in coincidence after one photon passes through a slit
mask. Strange as it may seem, the slit diffraction pattern is
observed on the detection screen of the other photon if one only
counts those pairs for which the first photon is detected by a
small-aperature fixed detector. For this EPR \cite{EPR} type
experiment, the data was interpreted in \cite{Strek} in terms of
entangled states. A second and similar experiment \cite{Pitt}
exhibited optical imaging via two photon coincidence counting.
This experiment was likewise analyzed in terms of entangled
states. In an earlier conference report \cite{dal98} we have shown
with detailed numerical calculations that "ghost" diffraction
patterns can be explained by trajectory correlations assuming an
initial position and velocity relationship for the particles.

In the present work, we give some detailed calculations of "ghost"
diffraction patterns and examine trajectory details in comparison
with those involved in deterministic explanations of ordinary
diffraction. The trajectory calculations are carried out
numerically using the trajectory-wave ordering interaction of
\cite{dal94,dal97}. The latter is a set of second order
differential equations for trajectories which invoke, via
solution, the integral method of generating random variants. We
begin with a brief description of this particle-wave interaction.

\section{TRAJECTORY-WAVE ORDERING INTERACTION (TWOI)}

    The TWOI studied earlier by the author, made use of a general
    line parameter \cite{dal94,dal97}. The following equations represent a version
     with time as the independent variable.
 \begin{equation}\label{T01}
\frac{d\dot{x}_i}{dt}-\frac{du_i}{dt}=-(\dot{x}_i-u_i)F
\end{equation}
In these equation the $x_i$  represent the particle coordinates,
and $\dot{x}_i=dx_i/dt$. The $u_i$ are components of a field
ray-velocity and the function $F$ is a real positive function of
field components with units of inverse time. The field affects the
trajectories only through the $u_i$ and function $F$. The TWOI
equations can be implemented with different choices for the $u_i$
and function $F$, facilitating the study of different trajectory
models. Specific expressions for $u_i$ and $F$ used in the
diffraction calculations here are discussed in the following
section. Modifications to impose metric constraints are discussed
in \cite{dal94,dal97}.

The above set of trajectory equations has in general two important
properties. First, the  $\dot{x}_i$ approach the $u_i$ as a limit.
Second, in this limit $(\dot{x}_i\rightarrow u_i)$ , ( named the
"attractor limit" in \cite{dal94} ) the large number distribution
of the particles approaches the function $F$ provided the initial
particle phase-space values are chosen uniformly randomly over a
large but not unique volume. The attractor limit is approached
closer and closer as time increases. For practical calculations,
one must set some criteria for "sufficiently close". These feature
can be more easily seen with a simple one-dimensional case with
$u_i=0$ . We directly integrate the following equation.
 \begin{equation}\label{T02}
\frac{d\dot{x}}{dt}=-(\dot{x})F
\end{equation}

This gives the following integral relations
\begin{equation}\label{T03}
\dot{x}(t)=\dot{x}(0)-\int^t_0F\dot{x}(t')dt'
\end{equation}
\begin{equation}\label{T04}
\dot{x}(t)=\dot{x}(0)-\int^x_{x_0}Fdx'
\end{equation}
If we numerically solve (\ref{T02}) until the attractor limit
$\dot{x}(t)\rightarrow0$ is sufficiently satisfied, we obtain the
following relation.
\begin{equation}\label{T05}
\int^x_{x_0}Fdx'\cong\dot{x}(0)
\end{equation}
Repeating this process many times with ($x(0),\dot{x}(0)$) chosen
uniformly randomly over a sufficient volume of phase space, we
generate a random variant specified by $F$. The quantity $Fdx$ is
proportional to the number of particles that reach the attractor
limit $\dot{x}(t)\rightarrow 0$ between $x$ and $x+dx$.  It is a
technique that indirectly invokes the integral method \cite{rubin}
to generate in the attractor limit a random variant specified by
$F$ . It re-arranges an initial distribution to match $F$.  This
interaction is viewed as one possible mechanism by which nature
could bring about the observed relation between particle
distributions and wave intensities.

To solve the trajectory equations we use both fourth and fifth
order Runge-Kutta methods \cite{dal94,Carna,Walt} with adaptive
step size techniques \cite{Kincade,Press}. These two methods are
used as checks against each other. The adaptive step technique
saves two to four orders of magnitude on net computation time. The
system of second-order differential equations is reduced to a
system of first-order differential equations following the method
of Butcher \cite{Butch64,Butch94}. Numerical details and a test
case are given in the appendix of \cite{dal94}.  Figure~\ref{f1}
presents a histogram of particle positions in the attractor limit
generated using (\ref{T02}) with $x(0)=-7$ ,
$F=0.3\exp(-.3|x|)\cos^2(2x)$ , and $\dot{x}(0)$ chosen uniformly
randomly from the range $(0,1.2)$. The "sufficiently close" to the
attractor limit criteria used was $|\dot{x}(t)|=10^{-5}
\max{|\dot{x}(0)|}$. The solid line in Fig.~\ref{f1} gives the
profile of $F$ scaled to match the particle histogram. This
distribution can also be generated with a wide variety of choices
of the initial phase space volume. Use of random or a variety of
amplitudes in $F$ (instead of $0.3$ ) together with a sufficiently
large initial phase-space volume will still generate the same
shape. The noise in the distribution is due to initializing with a
random number generator, the "sufficiently close" criteria and
numerical approximations.

  The second property of the TWOI equations is the alignment feature of the particle
  velocities with the wave ray velocity. From (\ref{T01}) we can see that the rate
   that the attractor limit is approached depends on
      the initial velocity difference  $(\dot{x}_i-u_i)$  and upon $F$.
      In Fig. ~\ref{f2} we give some representative trajectories with particles
       initialized at $Z=10\lambda$
       beyond a slit of width $10\lambda$. The particles are initialized in directions
       at $20^0$ intervals. One can see that the particles initially moving backwards can be
turned around and approximately aligned up with the wave ray
velocity in less than a wavelength.

The TWOI approach has some similiar features to, but some
conceptional differences from, the deterministic approach
initiated by Madelung  \cite{Madel} and de Broglie \cite{debro}
and
 re-considered later by Bohm \cite{bohm,Viger} and others \cite{hiley}
  for the Schrodinger
        equation.  In that work, the particle velocity $\dot{\vec{x}}$ is
        identified with the wave ray velocity so that the
        particles follow the wave flow tangent lines. We will refer to
        this case ( $\dot{\vec{x}}=\vec{u}$)
         as the attached
        case. This attached case is approached in the attractor limit for the
        TWOI.

        Critics of attached models
         quickly point out that the particle distribution must initially
         match the quadratic wave density, and that there is no mechanism
         to bring about the attached condition ($\dot{\vec{x}}=\vec{u}$).
          If initial conditions are met,
         the continuity equation guarantees that the distributions will
         continue to match. However, if tangent lines are initiated with a
         density that does not match the quadratic wave density, they will
         continue to differ.
         Interestingly, it was Bohm  \cite{bohm} who suggested using a force
          that depended on the difference ( $\dot{\vec{x}}-\vec{u}$ )
         to bring about
         the attached case as a limit. The TWOI can bring about this
         case as a limit, as well as generate in the particle distribution
         the wave intensity pattern
          if the latter is represented by $F$ in the TWOI equations.
          The TWOI is a somewhat more general approach in that it does
           not depend upon a continuity equation,  and $F$
          can represent any field intensity.  Likewise, the
          $\vec{u}$ can be chosen in different ways, e.g.,
          as a vector parallel to the Poynting vector as in
          \cite{prosser}, or as in the interesting work of Floyd
          \cite{Floyd}.

\section{SLIT DIFFRACTION}

    Slit diffraction has been observed for many types of particles,
    including photons, neutrons and atoms. Can slit diffraction
    be explained via calculating particle trajectories, one particle at
     a time?   One adamant view on this question is reflected in the
     well-known comment of Richard Feynman  " a phenomenon which is
     impossible, absolutely impossible, to explain in any classical
     way, and which has in it the heart of quantum mechanics " \cite{Feyn}.
     By treating the wave as a field and using the TWOI, the author
      has been able to obtain slit diffraction patterns, one particle
       at a time.  Here, we present a single, a double and a triple slit
       diffraction pattern in Figures ~\ref{f3}, ~\ref{f4}, and ~\ref{f5} respectively.
       These diffraction patterns are calculated using the following
       scalar wave equation $\nabla^2\Psi-k^2\Psi=0$  and standard
       Rayleigh-Sommerfield diffraction
       \cite{Good} with Neumann boundary conditions together with the following equation
       for $\vec{u}$.
\begin{equation}\label{T06}
\vec{u}=-\imath\frac{A}{2}(\Psi^{\dag}\vec{\nabla}\Psi-(\vec{\nabla}\Psi^{\dag})\Psi)/(\Psi^\dag\Psi)
\end{equation}

        For light diffraction, we use $A=c/k$  where  $c$ is the speed of light.
        With (\ref{T06}) we are taking $\vec{u}$ proportional to
        the gradient of the phase of the complex field, as in the
        work of  Madelung  \cite{Madel} and de Broglie
        \cite{debro}. The scalar field, together with its first and second partial
        derivatives needed in (\ref{T01}), are calculated numerically for each
        point along each trajectory.  Because an adjustable step technique
        is used, the exact position where the trajectory crosses a screen
        is determined by using a fifth order Lagrange interpolation
        polynomial \cite{Henrica}.

        In each figure, the wave is calculated assuming a point
        source $500\lambda$
        in front of the slits and a detector screen at $1000\lambda$ beyond the slits.
        For the single slit case shown in Fig.~\ref{f3}, the slit width
        is $8\lambda$.
        In the double slit case shown in Fig.~\ref{f4} , the slit width
        is $5\lambda$
         and the center-to-center slit separation is $25\lambda$ .  In the three slit
         case shown in Fig.~\ref{f5} , the slit width is $5\lambda$ , and the center-to-center
         slit separation is $10\lambda$.  The solid lines in these figures represents
         the wave intensity scaled to match the histograms.  From these figures,
         it is clear that slit diffraction patterns can be reproduced via
         deterministic trajectory calculations using the TWOI and treating
          the wave as a field.

It is instructive to view some typical trajectories.
Figure~\ref{f6} shows some typical trajectories for a three slit
case with slit width of $5\lambda$  and slit center to center
separation of $7\lambda$. These trajectories are near the
attractor limit at the slits, so that the trajectories closely
represent the wave flow tangent lines. Notice that these
trajectories are not straight lines and that the trajectories do
not cross each other. These feature are characteristic of tangent
line trajectories for waves involving the operator $\nabla^2$
\cite{Phili}.
\section{TWO-PHOTON GHOST DIFFRACTION}

 Two-photon correlations
involving slit diffraction have been measured using photon pairs
made in a parametric down conversion process \cite{Strek}.
Figure~\ref{f7}a illustrates the experimental setup.  The
"unfolded" sketch is illustrated in Fig.~\ref{f7}b. These figures
and notation follow Fig. 5 of Ref. \cite{Strek} with some changes
that feature the trajectory calculations here. The two
polarization components are separated by a polarizing beam
splitter into the "signal" beam and the "idler" beam as shown. The
signal beam passes through a slit pattern onto a screen about a
meter behind the slits.  No slit pattern is placed in the idler
beam.   A small aperture detector $D_1$ is placed at a fixed
position on the signal screen and a small aperture detector $D_2$
is used to scan the idler screen. Coincident two-photon
measurements are made for signal-idler photon pairs if a signal
photon enters the fixed detector $D_1$. After many pairs have been
measured, the accumulated distribution on the idler screen clearly
exhibits a slit diffraction pattern \cite{Strek}. This is
remarkable because no slit mask exist in the idler beam.

The author considers this to be one of the most intriguing
experimental results of modern physics. It would appear that what
happens to the signal photon is some how "known" by the idler
photon.  A two-photon entangled state explanation of the data
distributions was given in \cite{Strek}. Can this type of
experiment be explained with a deterministic interactive
particle-field model? To investigate this problem, we consider a
very simple model whose features are best illustrated by the
unfolded version sketch in Fig.~\ref{f7} b.  In this figure, as in
\cite{Strek} , the distance $Z_2$ is the distance from the slits
back through the beam splitter $BS$ to the crystal and then back
along the idler path to the idler screen. The distance $Z_0$ is
the distance from the crystal to the beam splitter and on to the
slits.  $Z_1$ is the distance from the slits to the fixed detector
$D_1$.

 We describe
this system using our interacting particle-field model. The
\textit{o}-ray and \textit{e}-ray polarization components satisfy
the wave equations
 $\nabla^2E_o-k^2E_o=0$
and  $\nabla^2E_e-k^2E_e=0$, and for both components we assume
spherical wave solutions with sources centered at some point S in
the crystal.  The \textit{e}-ray is diffracted by the slits and
the \textit{o}-ray continues to the idler screen.  We likewise
assume that together with the spherical wave, two particles are
initialized at point S in the crystal with equal and opposite
velocity vectors \cite{Strek}. For the \textit{e}-ray we calculate
the diffracted wave as in the slit diffraction described above,
but here the source point S is generally not on the optical axis.

To calculate the particle trajectories we use Eqs. (\ref{T01}) and
(\ref{T06}) with $A=c/k$, and $\Psi=E_o$  or $\Psi=E_e$.  Our
model crystal source is very thin in the $Z$ direction and for
each pair of particles, we choose the $Y$ position of the source
point S in the crystal uniformly randomly from a range of
$(-W,W)$. The initial angular direction of the signal particle is
likewise chosen uniformly randomly. However, the initial velocity
direction of the idler particle is always fixed as the negative of
the initial velocity direction of the signal particle (as viewed
in the unfolded picture of Fig.~\ref{f7}b). This and the same
initial position represent the initial trajectory coupling that
gives rise to the "ghost" diffraction in this model. We first
calculate the trajectory of the signal particle, and if it
terminates in the aperture of detector $D_1$, we calculate the
trajectory of the idler particle and record where it hits the
idler screen. With this coincidence restriction, the accumulated
histogram of idler particles on the idler screen is the "ghost"
pattern in the language of \cite{Strek}. An initial report by the
author on this type of two-photon "ghost" diffraction calculations
for a single and double slit case appears in a recent conference
proceedings \cite{dal98}.

 We show in Fig.~\ref{f8} and Fig.~\ref{f9} the
"ghost" diffraction patterns calculated with this simple model for
cases of one and three slits respectively.  For both cases we used
$Z_2=7000\lambda$, $Z_0=1000\lambda$, and $Z_1=500\lambda$. For
the single slit case, shown in  Fig.~\ref{f8} we used a slit width
of $15\lambda$, a crystal half width of $W=155\lambda$ , and a
$6\lambda$ aperture for the signal detector $D_1$. For the three
slit case shown in Fig.~\ref{f8}, we used a slit width of
$5\lambda$, a slit center-to-center separation of $7\lambda$, a
crystal half width of $W=215\lambda$, and a
 $10\lambda$ aperture for the detector $D_1$.  It was necessary to
 calculate many signal trajectories because very few terminated on the
 narrow detector aperture of $D_1$.

    The distinct recognizable diffraction patterns exhibited in
      Fig.~\ref{f8} and Fig.~\ref{f9} clearly indicate that two-photon "ghost"
     diffraction can be explained via deterministic trajectory
     calculations using the interacting particle-field model.
     This two-photon correlation phenomenon is a result of pattern
     selection via correlated trajectory selection.  The trajectory
     correlations are entirely due to initial conditions.
     In Fig.~\ref{f8} and Fig.~\ref{f9} the solid line represents the diffraction
     pattern on the idler screen that results as if one had a point
      source at the signal detector on the signal screen. The common
      features of the solid line and the histogram are due to the common
      geometry in the almost Fraunhofer limit.  However, the patterns
       differ slightly from each other. This difference can be reasonably explained by
       the fact that the signal detector $D_1$ has a finite size.
       It is conceptually incorrect to think that the "ghost" diffraction is
       the result of a "collapsed probability" signal at $D_1$ propagating
       backwards to the idler screen.

    In the conventional slit diffraction of
     Fig.~\ref{f3}, Fig.~\ref{f4}, and Fig.~\ref{f5} the wave
     used for each particle had a common point source.  By sharp contrast,
      each particle in the "ghost" diffraction generally is influenced by
      a different wave since the source positions in the crystal have a wide range
      $(-W,W)$ . A random selection of signal trajectories just beyond the slits is shown
        in Fig.~\ref{f10} for the three slit case.   One can see that the trajectories
        cross each other and appear to have no definite pattern.  In fact, if one
        accumulates a histogram on the signal screen, one finds a broad smear with
        no definite pattern.  This is expected because each wave generally starts
        from a different source points in the crystal.  However, it is in
        this very
        situation that we obtain the "ghost" diffraction pattern on the idler
         screen for that selection of signal trajectories that terminate in the
         fixed detector $D_1$.  This feature is consistent with the observations
         in \cite{Strek}, in which the "ghost" patterns were obtained using rather
         divergent beams. By shifting the position of the detector $D_1$ we
         obtain a shifted "ghost" pattern on the idler screen. This corresponds
          to selecting an entirely different set of signal-idler trajectory pairs.
          The authors in \cite{Strek} mentions that this interesting feature was
          observed in the experiment.

\section{SUMMARY}

We have made "ghost" diffraction calculations using a trajectory
model with two key features. The first feature is the use of the
TWOI equations to describe the trajectories. Through the TWOI, the
diffracted field influences the distribution of the trajectory
particles, one trajectory at a time. As a physical model both
particle and field must be taken together.  Here, the TWOI used to
calculate the particle-field interaction is much closer to that of
classical physics in which the particle is not "attached" to the
field. This key freedom, not found in the "attached" models
initiated by Madlung \cite{Madel} and de Broglie \cite{debro} many
years ago, facilitates the random variant generator inherent in
the structure of the TWOI to generate the wave intensity patterns
in the ensemble distribution.

The second key feature is the initial velocity correlation of the
photons in the down conversion crystal, as well as the same
initial position. As the calculations clearly demonstrate, this
initial coupling together with the TWOI is sufficient to give rise
to the "Ghost" diffraction patterns. This is a simple classical
trajectory model of this photon-photon experiment. Its success
clearly lends support to a realistic and deterministic description
of nature.

\begin{figure}[htbp]
  \vspace*{19.00cm}
  \hspace*{5.0cm}
 \includegraphics{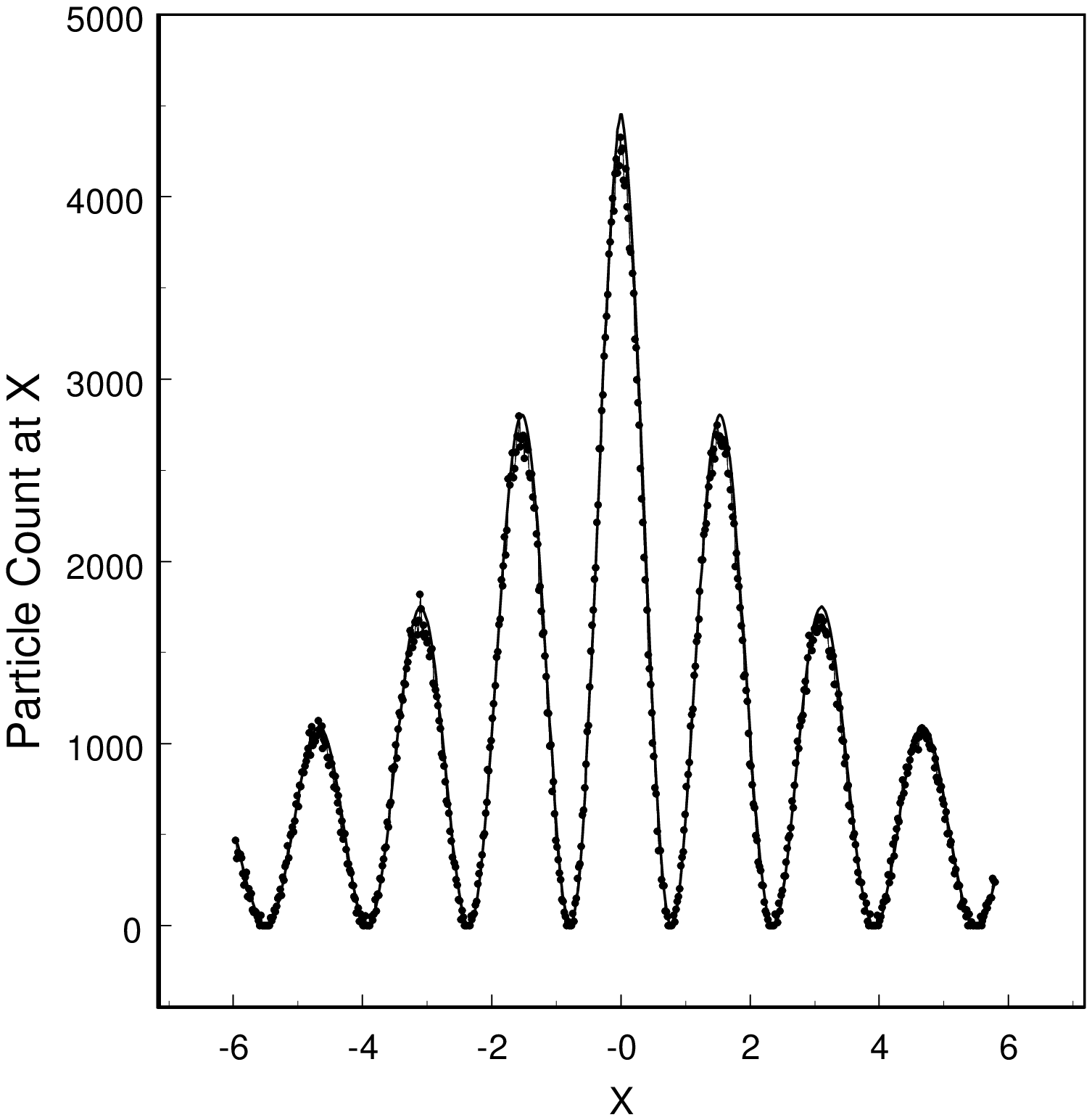}
 \vskip -1.8in
    \caption{A one dimensional histogram of particle count versus the $X$ position in
    the \\ attractor limit. The solid line is the shape of the function $F$ scaled
     to approximately \\ match the histogram.}
    \label{f1}
\end{figure}
\begin{figure}[htbp]
  \vspace*{19.00cm}
  \hspace*{5.0cm}
 \includegraphics{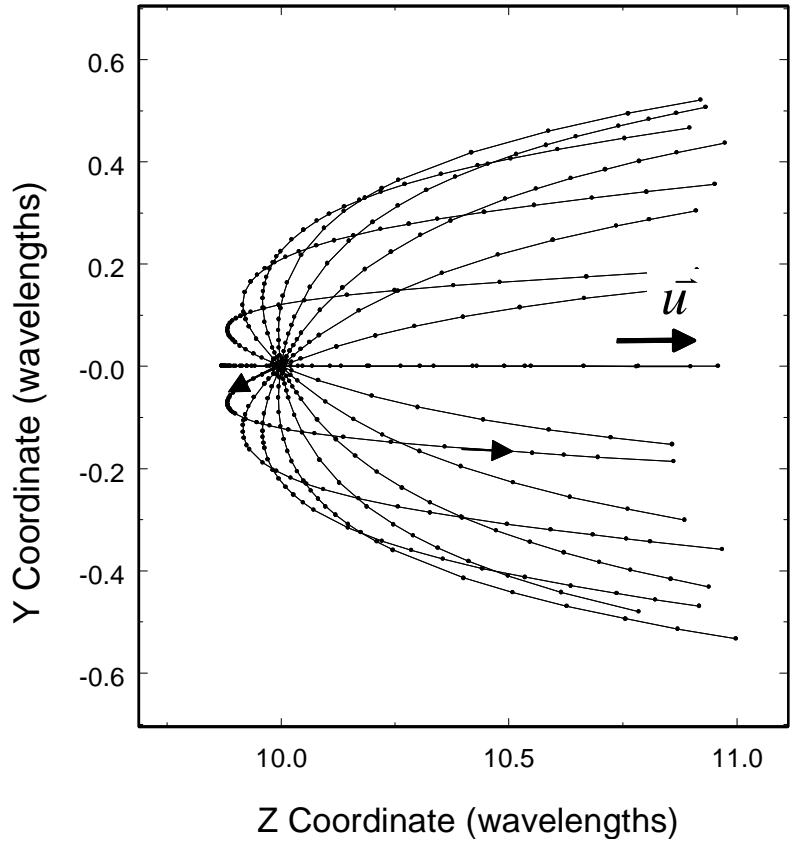}
 \vskip -2.in
    \caption{Typical trajectories illustrating how the TWOI can bring
    about the \\ attractor limit $(\dot{x}_i\rightarrow u_i)$.  The
    trajectories are initialized at the same point in \\ directions
    at $20^0$ intervals.}
    \label{f2}
\end{figure}
\begin{figure}[htbp]
  \vspace*{19.00cm}
  \hspace*{5.0cm}
 \includegraphics{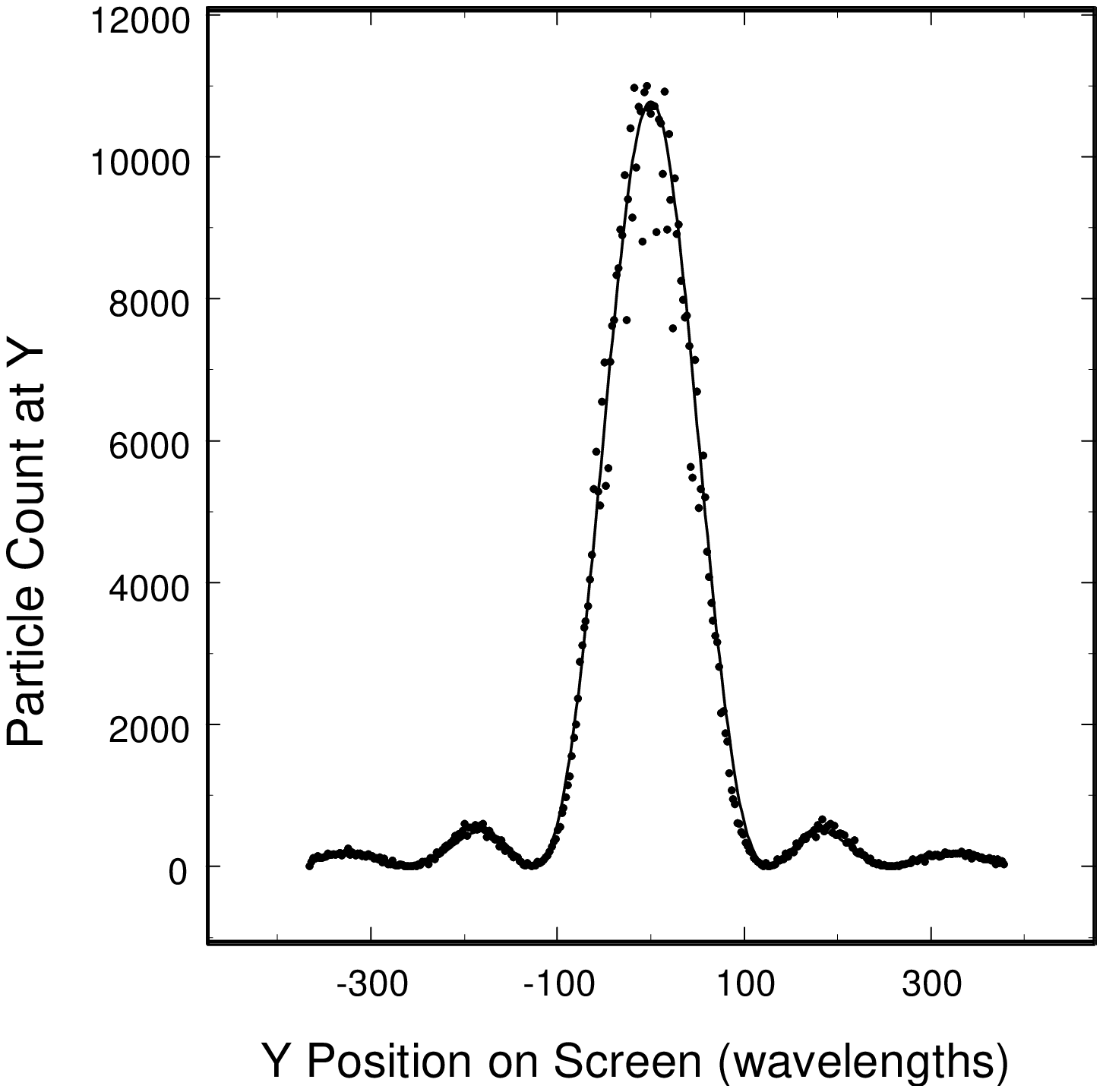}
 \vskip -2.in
    \caption{Histogram of particle count versus the Y position on the
detector screen \\ for single slit diffraction. The trajectories
of the particles were calculated one \\ at a time. The solid line
represents the wave intensity curve scaled to approximately match
\\ the histogram.}
    \label{f3}
\end{figure}
\begin{figure}[htbp]
  \vspace*{19.00cm}
  \hspace*{5.0cm}
 \includegraphics{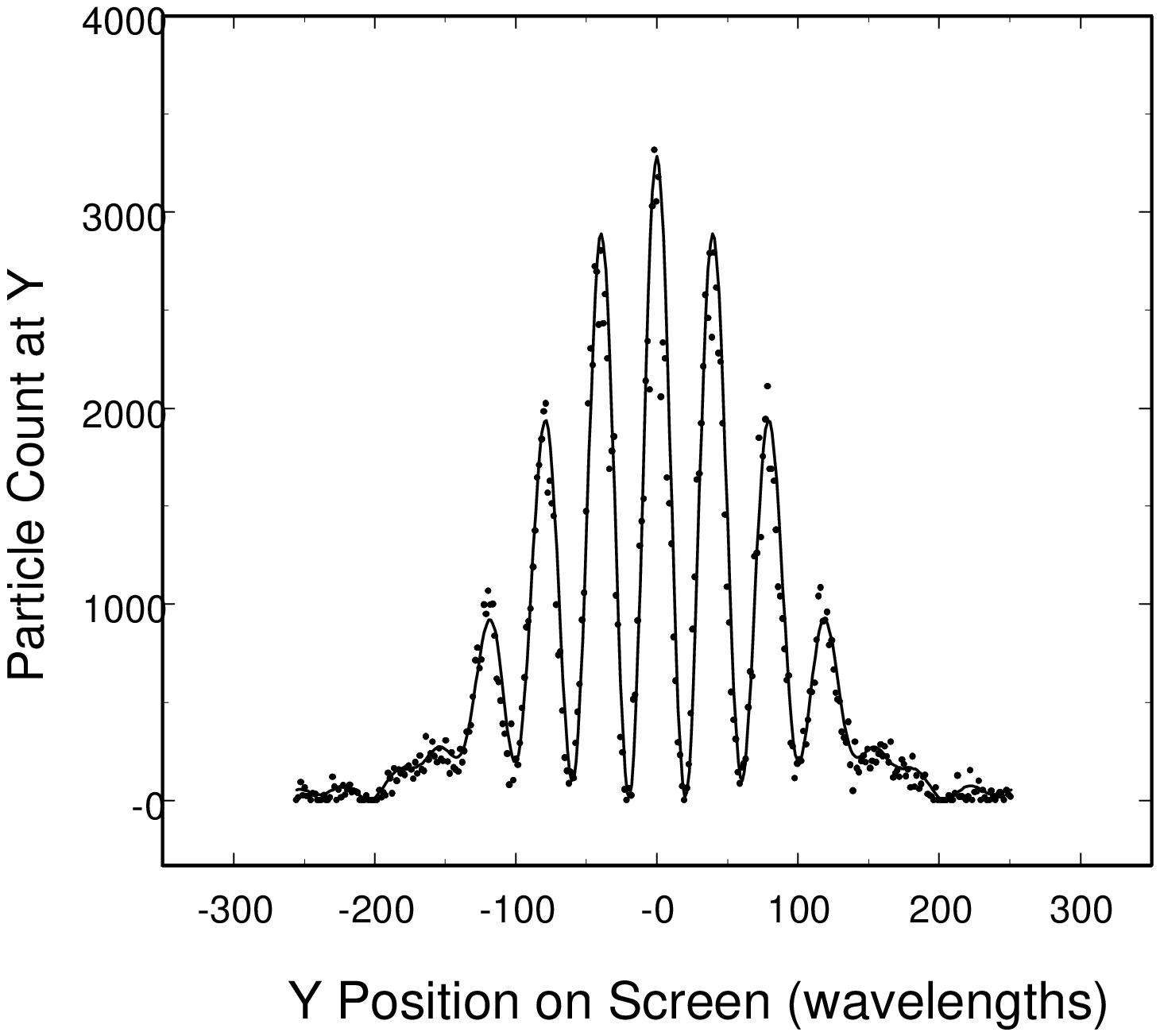}
 \vskip -2.1in
    \caption{Histogram of particle count versus the Y position on the
detector screen \\ for double slit diffraction. The trajectories
of the particles were calculated one at \\ a time. The solid line
represents the wave intensity curve scaled to approximately match
\\ the histogram.}
    \label{f4}
\end{figure}
\begin{figure}[htbp]
  \vspace*{19.00cm}
  \hspace*{5.0cm}
 \includegraphics{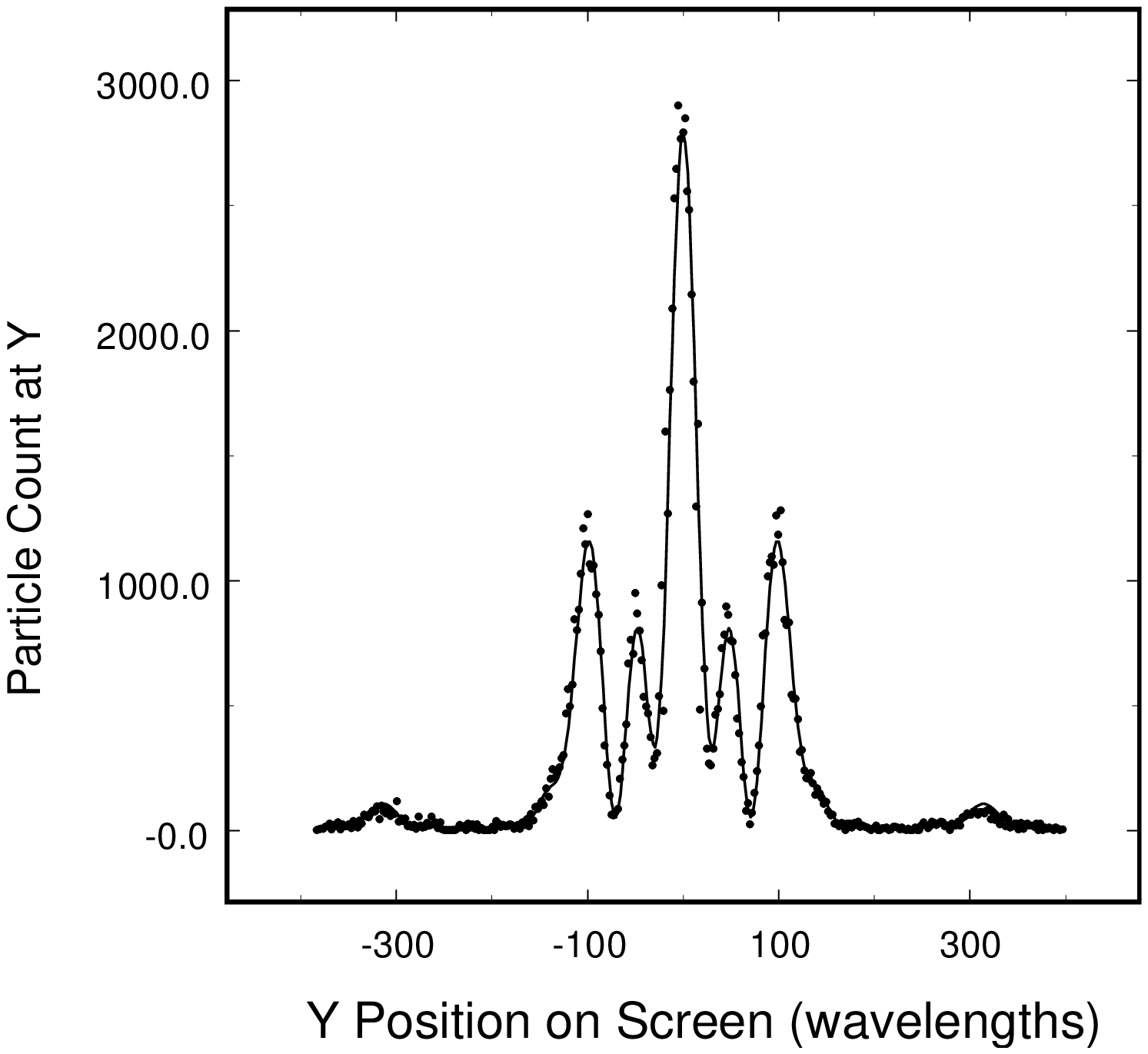}
 \vskip -2.2in
    \caption{Histogram of particle count versus the Y position on the
detector screen \\ for three-slit diffraction. The trajectories of
the particles were calculated one at \\ a time. The solid line
represents the wave intensity curve scaled to approximately match
\\ the histogram.}
    \label{f5}
\end{figure}
\begin{figure}[htbp]
  \vspace*{19.00cm}
  \hspace*{5.0cm}
 \includegraphics{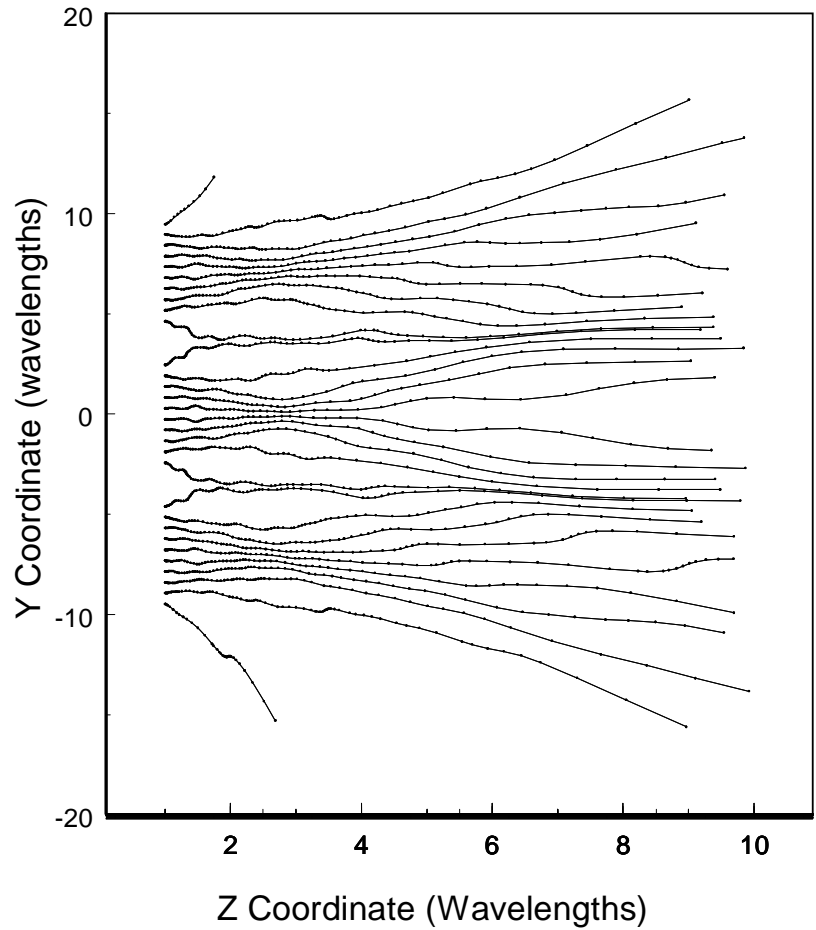}
 \vskip -1.5in
    \caption{A selection of typical trajectories in the
    attractor limit in the very \\ near field region for
    three-slit diffraction. The dots on the lines represent
     the \\ adjustable step positions. The trajectories are not
     straight lines and do not \\ cross each other.}
    \label{f6}
\end{figure}
\begin{figure}[htbp]
  \vspace*{19.00cm}
  \hspace*{5.0cm}
 \includegraphics{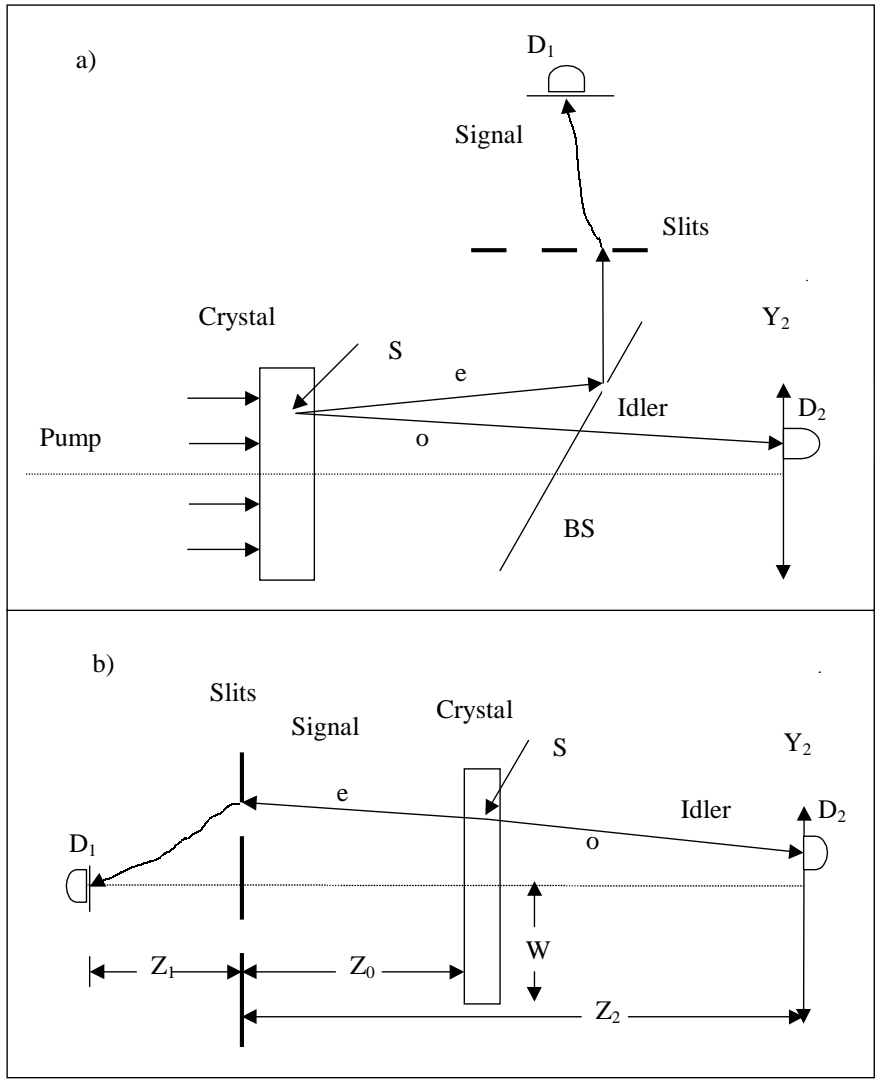}
 \vskip -.5in
    \caption[]{Schematic (a) illustrates the experimental setup and calculation
model for \\ "ghost" diffraction, and schematic (b) illustrates
the unfolded view following Fig. $5$ of  \cite{Strek}. \\
Detectors D1 and D2 measure the signal and idler photons in
coincidence. D1 is fixed \\ and D2 is scanned. The diffraction
pattern is observed on the idler screen whereas the \\ signal
screen is behind the slits.}
    \label{f7}
\end{figure}
\begin{figure}[htbp]
  \vspace*{19.00cm}
  \hspace*{5.0cm}
 \includegraphics{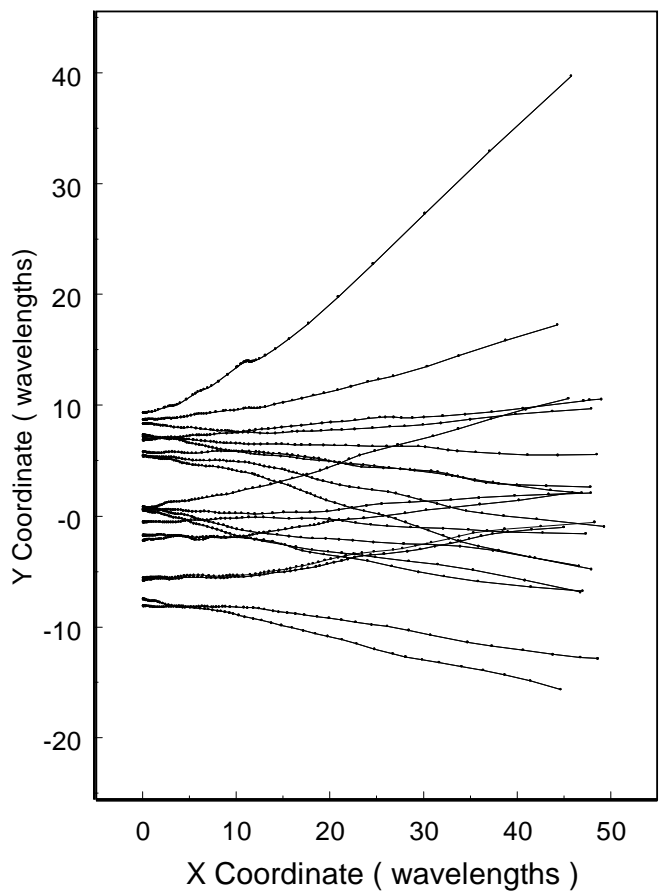}
 \vskip -1.8in
    \caption{A random selection of signal trajectories just
beyond the three-slit mask. \\ Even though the particle is each
case is near the attractor limit, the trajectories can \\ cross
each other because each is influenced by a wave from a different
point source \\ location in the crystal. The "ghost" diffraction
pattern arises from those idler \\ trajectories paired with just
the very small fraction of signal trajectories that \\ terminate
on the small detector $D_1$.}
    \label{f8}
\end{figure}
\begin{figure}[htbp]
  \vspace*{19.00cm}
  \hspace*{5.0cm}
 \includegraphics{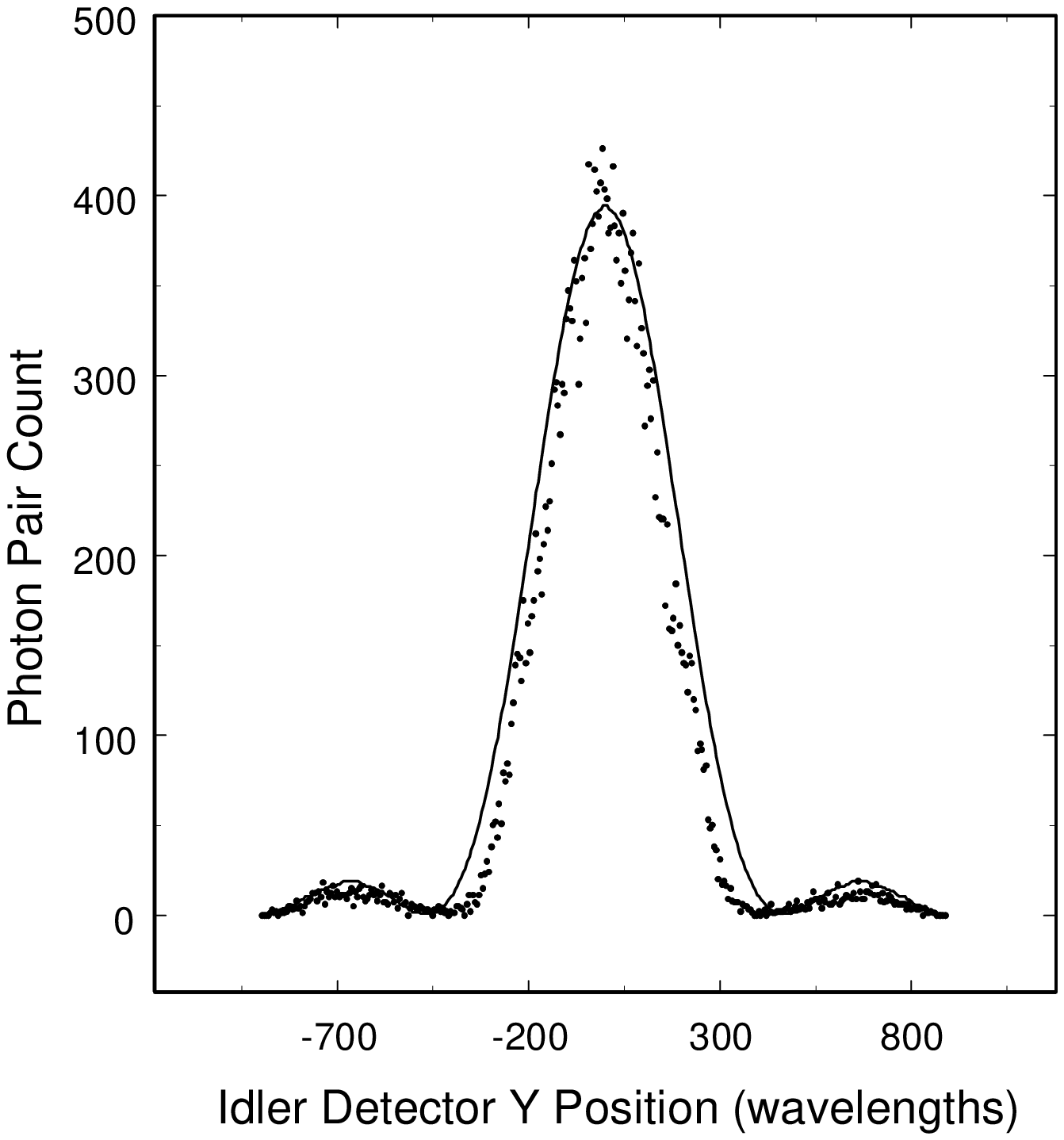}
 \vskip -1.9in
    \caption{A single slit "ghost" diffraction pattern.  Histogram of
photon pair count \\ versus Y position on the idler screen.  The
correlated signal and idler trajectory \\ pairs are calculated one
at a time.  The solid line represents a diffraction \\ curve
calculated on the idler screen as though one had a point source \\
at the signal detector.}
    \label{f9}
\end{figure}
\begin{figure}[htbp]
  \vspace*{19.00cm}
  \hspace*{5.0cm}
 \includegraphics{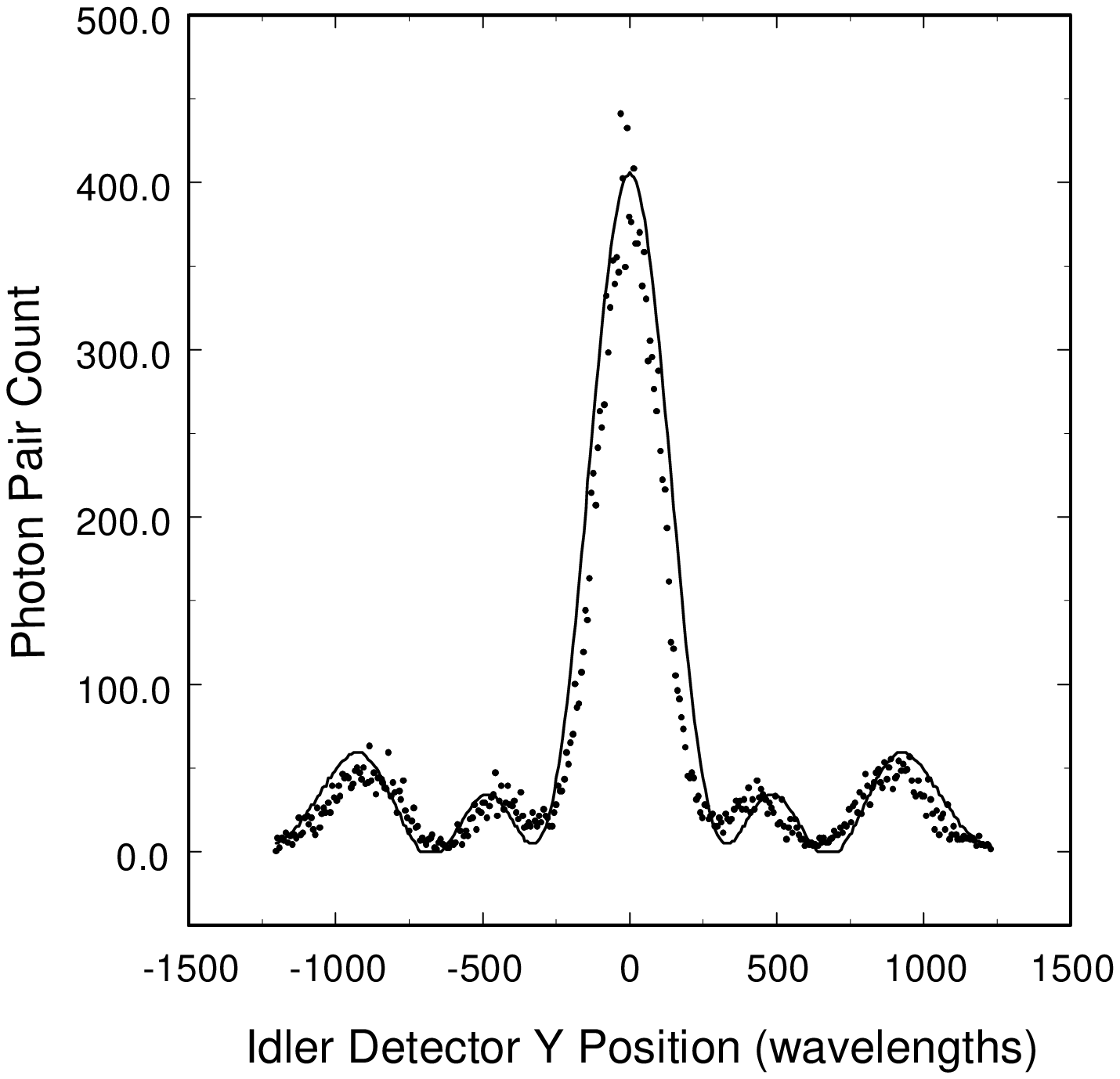}
 \vskip -2.2in
    \caption{A three slit "ghost" diffraction pattern. Histogram of photon
pair \\ count versus Y position on the idler screen.  The
correlated signal and idler \\ trajectory pairs are calculated one
at a time. The solid line represents a diffraction \\ curve
calculated on the idler screen as though one had a point
\\ source at the signal detector.}
    \label{f10}
\end{figure}

\begin{thebibliography}{}


\bibitem{Strek}
 Strekalov D.V., Sergienko A. V., Klyshko D. N., and Shih Y. H.
 (1993), Phys. Rev. Lett, 74 (18) 3600.
\bibitem{EPR}
 Einstein A., Podolsky B., and Rosen N. (1935), Phys. Rev. 47, 777.
\bibitem{Pitt}
  Pittman T. B.,  Shih Y. H., Strakalov D. V., and Sergienko A.
  V. (1995), Phys. Rev. A 52,3429.
\bibitem{dal98}
  Dalton B. J. (1998), in Causality and Locality in Modern Physics,
  Eds. G. Hunter, S. Jeffers, and J-P. Vigier, (Kluwer-Academic
  Publishers, Dordrecht, The Netherlands), 463.
\bibitem{dal94}
 Dalton B. J. (1994), Deterministic Explanation of Quantum Mechanics
 Based on a New Trajectory-Wave Ordering Interaction, North Star
 Press of St. Cloud Inc., St Cloud, Minnesota.
\bibitem{dal97}
 Dalton B. J. (1997), in The Present Status of the Quantum Nature of
 Light, Eds. S. Jeffers, S. Roy, J-P. Vigier, and G.
 Hunter,(Kluwer-Academic Publishers, Dordrecht, The Netherlands), 235.
\bibitem{rubin}
 Rubinstein R. Y. (1981), Simulation and The Monte Carlo Method,
 Wiley and Sons, N. Y.
\bibitem{Carna}
  Carnahan B.,Luther H. A. , Wilkes J. O. (1969), Applied
 Numerical Methods, John Wiley and Sons, N. Y., 366.
\bibitem{Walt}
 Walters J.(1996), Commun. Ass. Comp Mach. 9 293.
\bibitem{Kincade}
 Kincaid D., Cheney W. (1990), Numerical Analysis, Brooks/Cole Publishing,
 Pacific Grove, California, 502 .
\bibitem{Press}
  Press W. H.,Flannery B. P.,
 Teukolsky S. A., and Vetterling W. T.(1989), Numerical Recipes,
 The Art of Scientific Computing, Cambridge University Press,
 Cambridge CB2 IRP.
\bibitem{Butch64}
 Butcher J. C. (1964), J. Austral. Math. Soc. 4, 179.
\bibitem{Butch94}
 Butcher J. C. (1994), Computers in Physics, 8, (4) 411.
\bibitem{Madel}
Madelung E. (1926), Zeits. Phys.,40 332.
\bibitem{debro}
de Broglie L. (1927), Compt. Rendus., 184, 273; 185, 380.
\bibitem{bohm}
 Bohm D. (1952), Phys. Rev. 85 166; (1953), 89, 458.
\bibitem{Viger}
 Bohm D. and Vigier J. V. (1954), Phys. Rev., 96, 208.
\bibitem{hiley}
 Bohm D. and Hiley B. J. (1993),  The Undivided Universe, Routhledge,
 London and New York.  Many references to and summaries of earlier
 work can be found in this text.
 \bibitem{prosser}
 Prosser R. D., Jeffers S. and Sesroaches, J. (1997), in The Present
  Status of the Quantum Nature of
 Light, Eds. S. Jeffers, S. Roy, J-P. Vigier, and G.
 Hunter,(Kluwer-Academic Publishers, Dordrecht, The Netherlands), 151,
 This report list references to several related earlier studies.
\bibitem{Floyd}
 Floyd E. R. (1999), Int. J. Mod. Phys. A, 14, 111; See earlier
 references in this article.
\bibitem{Feyn}
 Feynman R. P., Leighton R. B. , Sands M.(1963), The Feynman
 Lectures on Physics, Vol. 1, Addison-Wesley, New York, 37.
\bibitem{Good}
 Goodman J. W.(1996), Introduction to Fourier Optics, McGraw-Hill,
 New York, 50.
\bibitem{Henrica}
  Henrica P. (1964), Elements of Numerical Analysis, John Wiley
 and Sons, Inc., New York, Ch. 9.
\bibitem{Phili}
  Philippidis C., Dewdney C., and Hiley B. J. (1979), Nuovo Cimento B 52,
  15.


\end{thebibliography}
\end{document}